\documentclass[aps,pre,preprint,superscriptaddress,floatfix,groupedaddress]{revtex4-1}
\usepackage{amsmath,amsfonts,amssymb}
\usepackage{graphicx,color}
\usepackage{bm}

\begin{document}

\title{Noise-enhanced chaos in a weakly coupled GaAs/(Al,Ga)As superlattice}
\author{Zhizhen Yin}
\author{Helun Song}
\author{Yaohui Zhang}
\affiliation{Key Laboratory of Nano devices and Applications, Suzhou Institute of Nano-tech and Nano-bionics,
Chinese Academy of Sciences, Suzhou 215125, China}

\author{Miguel Ruiz-Garc\'{\i}a}
\author{Manuel Carretero}
\author{Luis L. Bonilla}
\affiliation{Gregorio Mill\'an Institute, Fluid Dynamics, Nanoscience
and Industrial Mathematics, Universidad Carlos III de Madrid, 28911
Legan\'es, Spain}

\author{Klaus Biermann}
\author{Holger T. Grahn}
\affiliation{Paul-Drude-Institut f\"{u}r Festk\"{o}rperelektronik, Leibniz-Institut
im Forschungsverbund Berlin e.~V., Hausvogteiplatz 5--7, 10117 Berlin, Germany}

\date{\today}
\begin{abstract}
Noise-enhanced chaos in a doped, weakly coupled GaAs/Al$_{0.45}$Ga$_{0.55}$As
superlattice has been observed at room temperature in experiments as well as
in the results of the simulation of nonlinear transport based on a discrete
tunneling model. When external noise is added, both the measured and simulated
current-versus-time traces contain irregularly spaced spikes for particular
applied voltages, which separate a regime of periodic current oscillations from
a region of no current oscillations at all. In the voltage region without current
oscillations, the electric field profile consist of a low-field domain near the
emitter contact separated by a domain wall consisting of a charge accumulation layer
from a high-field regime closer to the collector contact. With increasing noise amplitude,
spontaneous chaotic current oscillations appear over a wider bias voltage range.
For these bias voltages, the domain boundary between the two electric-field
domains becomes unstable, and very small current or voltage fluctuations
can trigger the domain boundary to move toward the collector and induce
chaotic current spikes. The experimentally observed features are qualitatively
very well reproduced by the simulations. Increased noise can consequently
enhance chaotic current oscillations in semiconductor superlattices.
\end{abstract}

\maketitle

\section{Introduction}
With the continuous advancement of nonlinear science and the theory of random
dynamical system, we have become aware of the fact that noise should not always be
considered as a disturbing factor. Actually, a small amount of noise may enhance
the dynamics of a system so that it becomes better defined and controllable.
Constructive effects of noise in nonlinear systems have been investigated
extensively in the context of stochastic resonances and coherence resonances.
By stochastic resonance, noise can optimize the response of a system to an external
signal and induce stochastic phase synchronization to the external force~\cite{wie95,gam98}.
For a coherence resonance, pure noise without an external signal can generate the most coherent
motion in the system, as has been mainly observed in excitable systems~\cite{hu93,zho01}.

A doped, weakly coupled semiconductor superlattice (SSL) represents an almost ideal spatially one-dimensional
nonlinear dynamical system with a large number of degrees of freedom, the nonlinearity
of which is due to sequential resonant tunneling between adjacent quantum wells.
Fluctuations of the layer thicknesses, electron density, energy levels, and inter-well
coupling transform a weakly coupled SSL into a complex nonlinear system, in which the
electron transport is strongly dissipative. A great richness of nonlinear transport
behavior has been observed in weakly coupled SSLs, including the formation of stationary
electric-field domains, periodic as well as quasi-period current self-oscillations, and
even driven as well as undriven chaos~\cite{bon02,bon05}. The oscillatory behavior is
attributed to the localized, oscillatory motion of the domain boundary, which separates
the high from the low electric-field domain. Only very recently, spontaneous
chaotic~\cite{hua12} and quasi-periodic~\cite{hua13} current self-oscillations were observed at room
temperature in GaAs/(Al,Ga)As SLs using an Al content of $45$\%, which results in the
largest direct barrier for this materials system.

Both theoretical and experimental results have proven that noise can affect
the charge transport in weakly coupled GaAs/AlAs SLs~\cite{hiz06,bom12}.
A noise-enhanced coherence resonance has been observed in a weakly coupled GaAs/AlAs SL at
$77$~K~\cite{hua14}. The main reason for the existence of such a coherence
resonance in such a SL is related to the interaction between the noise and the two
oscillation modes existing in weakly coupled SSL oscillators such as the well-to-well hopping mode and the dipole-motion
mode. On this basis, numerical simulations have shown that noise enhances spontaneous
chaos at room temperature~\cite{alv14,bon16}.

In this paper, we report clear evidence for noise-enhanced spontaneous chaos in a doped,
weakly coupled GaAs/Al$_{0.45}$Ga$_{0.55}$As SL at room temperature. We find that, with
increasing noise amplitude, the spontaneous chaotic oscillations appear over a wider range of voltages.
The experimentally observed results are qualitatively very well validated by numerical simulations
based on a discrete resonant tunneling model that captures the main features of vertical transport
in doped, weakly coupled SSLs~\cite{alv14,bon16}. The paper is organized as follows. In the
next section, we describe the sample structure and measurement techniques followed by the experimental
results of the current-voltage characteristics, the current self-oscillations, and attractors extracted
from the experimental results. In Sec.~\ref{SecIII}, we present the discrete resonant-tunneling
model and the results of the numerical simulations.
Finally, we summarize the obtained results and conclude in Sec.~\ref{SecIV}.

\section{Experimental results \label{SecII}}
\subsection{Sample structure and measurement techniques}
\begin{figure}[!t]
\includegraphics[width=0.45\textwidth]{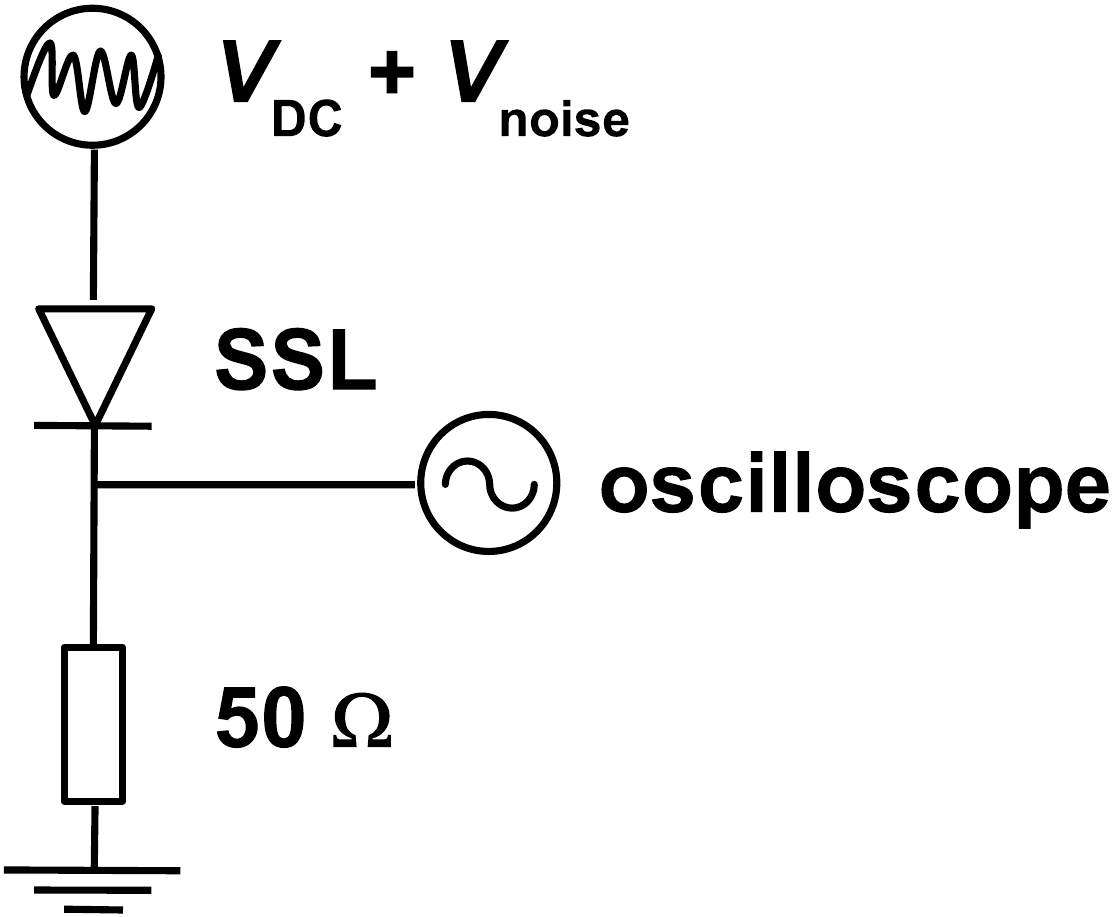}
\caption{Schematic of the experimental setup. The DC bias voltage $V_\text{DC}$
and the noise with amplitudes $V_\text{noise}$ are supplied
by a function generator. The current-time traces and the frequency spectra
are recorded with a $6$~GHz oscilloscope over a 50 $\Omega$ resistor.}
\label{Yin_Fig1}
\end{figure}

The sample consists of a doped, weakly coupled GaAs/(Al,Ga)As SL with $50$ periods, each period
consisting of a GaAs quantum well and an Al$_{0.45}$Ga$_{0.55}$As barrier. The central $3$~nm
of each $7$-nm-thick GaAs well are doped with Si at a density of $2$$\times$$10^{17}$~cm$^{-3}$.
The thickness of the Al$_{0.45}$Ga$_{0.55}$As barriers is $4$~nm, resulting in a rather weak coupling
between adjacent quantum wells. The SSL is sandwiched between two highly doped GaAs contact layers
forming an $n^+$-$n$-$n^+$ diode. For more details of the sample structure, see Ref.~\onlinecite{li15}.
After plasma etching and providing Ohmic contacts of AuGe/Ni/Au with $35/10/500$~nm using electron
beam evaporation followed by rapid thermal annealing at $420~^\circ$C, square mesas
with a side length of $30~\mu$m are investigated. Single devices are wire bonded for electrical measurements. All
experimental measurements are performed at room temperature. The DC bias voltage $V_\text{DC}$
and the noise with amplitudes $V_\text{noise}$ are applied using the function generator
Agilent 33220A. The current-time traces and the frequency spectra are recorded
with a $6$~GHz oscilloscope LeCroy Wavepro 760Zi-A. For the current-time traces,
the oscilloscope records voltage pulses, which are then converted in current values using the $I$-$V$ characteristics.
A schematic of the measurement circuit is shown in Fig.~\ref{Yin_Fig1},
which is the same as the one in Ref.~\onlinecite{hua14}.

\begin{figure}[!b]
\includegraphics[width=0.45\textwidth]{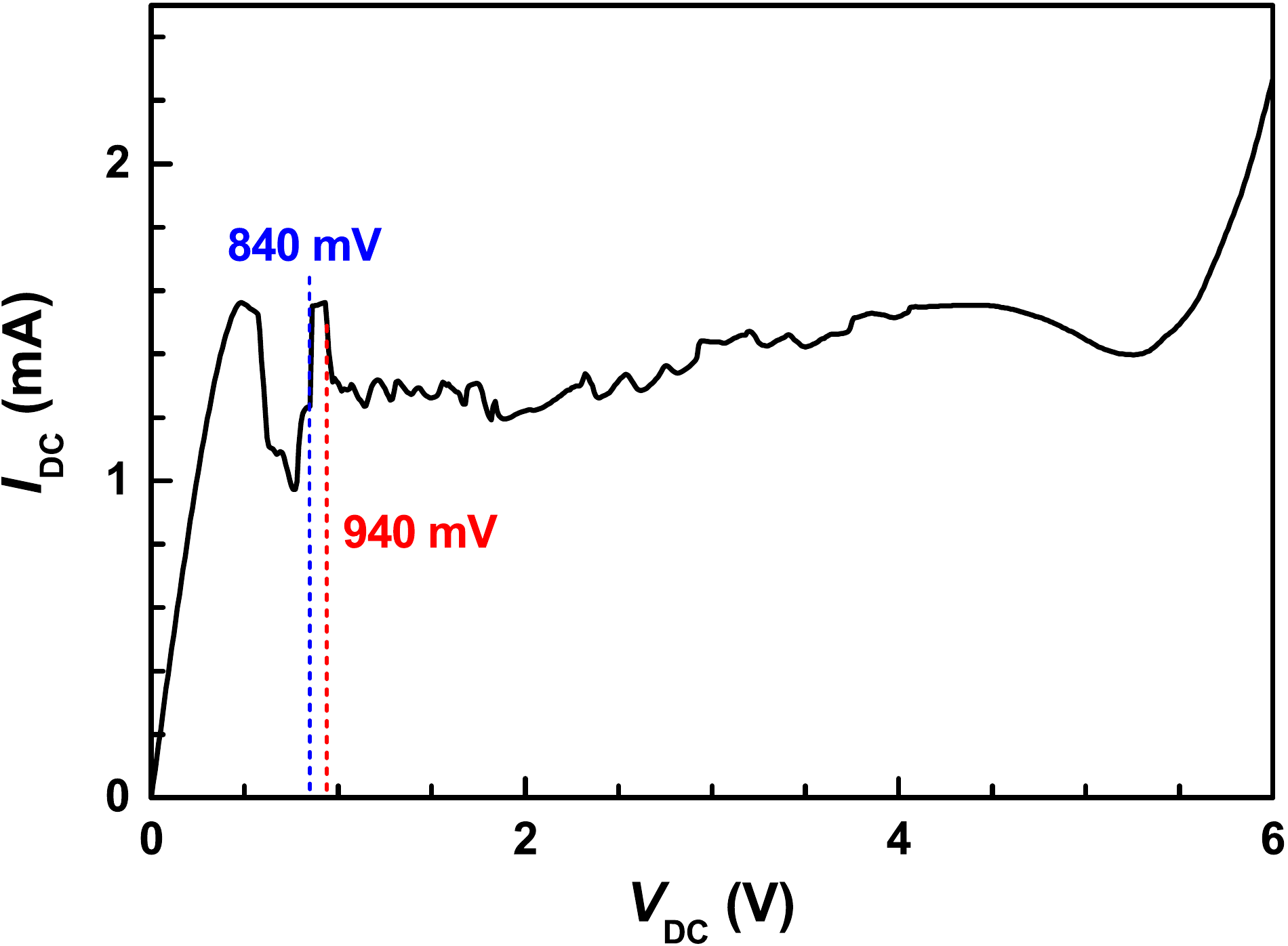}
\caption{Measured current $I_\text{DC}$ versus voltage $V_\text{DC}$ of the doped, weakly coupled
GaAs/Al$_{0.45}$Ga$_{0.55}$As SL at room temperature. At $V_\text{DC}=840$ and $940$~mV, the frequency
spectra are dominated by broad-band noise.}
\label{Yin_Fig2}
\end{figure}
\subsection{Current-voltage characteristics}
A typical current-voltage characteristic of the the doped, weakly coupled SSL
device is shown in Fig.~\ref{Yin_Fig2}. In the voltage range between $850$ and $860$~mV, the
DC current rapidly increases from $1.24$ to $1.55$~mA, followed by a current plateau
between $860$ and $930$~mV. In this voltage range, electric-field domains are
formed with a charge accumulation layer forming the domain boundary. As the voltage increases
further to $940$~mV, the current decreases to $1.50$~mA and then continues down to $1.31$~mA.

\subsection{Current self-oscillations and frequency spectra}
\begin{figure}[!b]
\includegraphics[width=0.45\textwidth]{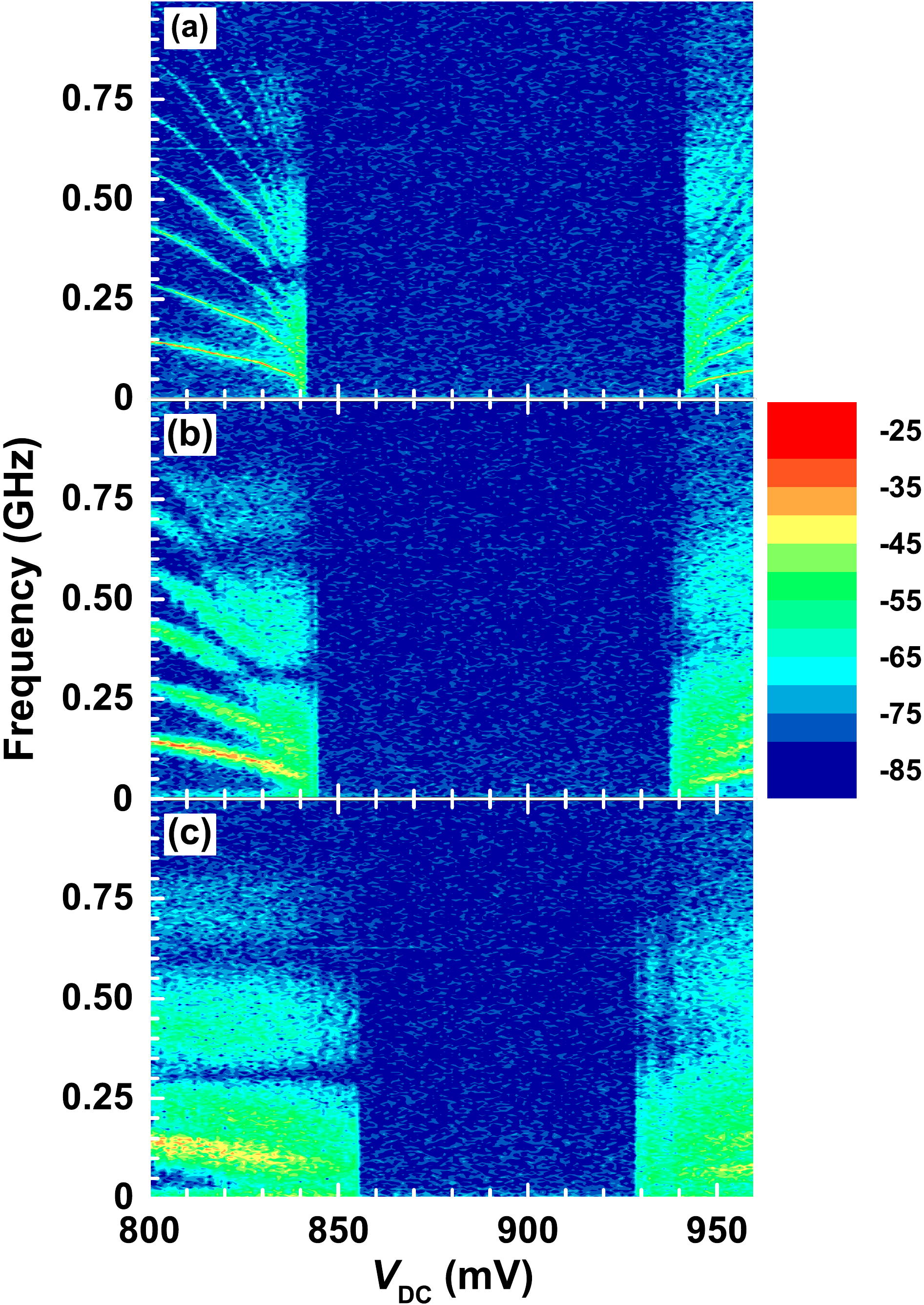}
\caption{Measured frequency spectra of the current versus $V_\text{DC}$
recorded between $800$ and $960$~mV (a) without noise, (b) with $V_\text{noise}=20$~mV,
and (c) with $V_\text{noise}=80$~mV.}
\label{Yin_Fig3}
\end{figure}

The frequency spectra displayed in Fig.~\ref{Yin_Fig3}(a) recorded between $V_\text{DC}=800$ and $960$~mV
without any external noise do not show any current self-oscillations for $V_\text{DC}$ values between
$840$ and $940$~mV. For $V_\text{DC}$ values outside this region, i.~e., smaller than $840$ or larger than $940$~mV,
the frequency spectra exhibit periodic oscillations with higher-order harmonics, and the fundamental
frequency depends on  $V_\text{DC}$. At $V_\text{DC}=840$ and $940$~mV, where the transitions between
current oscillations and no current oscillations occur, the frequency spectra are dominated by
an abrupt decrease to zero frequency. These transitions are accompanied by an abrupt change of the
current as shown in Fig.~\ref{Yin_Fig2}. This observation may indicate an infinite period collision
of the oscillatory attractor with a homoclinic orbit. Numerical simulations confirm this scenario
for the SSL model.

Figures~\ref{Yin_Fig3}(b) and \ref{Yin_Fig3}(c) show the frequency spectra for applying additional
external noise with amplitudes of $V_\text{noise}=20$ and $80$~mV, respectively.
For $V_\text{noise}=20$~mV in Fig.~\ref{Yin_Fig3}(b), the chaotic oscillations in the transition regions
appear over a larger voltage interval than without any external noise [cf. Fig.~\ref{Yin_Fig3}(a)].
The voltage region with no current oscillations becomes narrower. Increasing the noise amplitude to $80$~mV
results in a further shrinkage of the voltage region with no current oscillations, and the voltage range of
seemingly chaotic oscillations extends to $855$ and $929$~mV as shown in Fig.~\ref{Yin_Fig3}(c). At the same time,
the frequency spectra become broader with increasing noise amplitude. These experimental data clearly
demonstrate that noise can significantly enhance the chaos generated in doped, weakly coupled SSLs.
This was shown theoretically for the case of internal (shot and thermal) noise by the results of the
numerical simulations in Ref.~\onlinecite{alv14}. The present results show that chaos enhancement can
also be achieved more generally by adding appropriate external noise.

\begin{figure}[!b]
\includegraphics[width=0.45\textwidth]{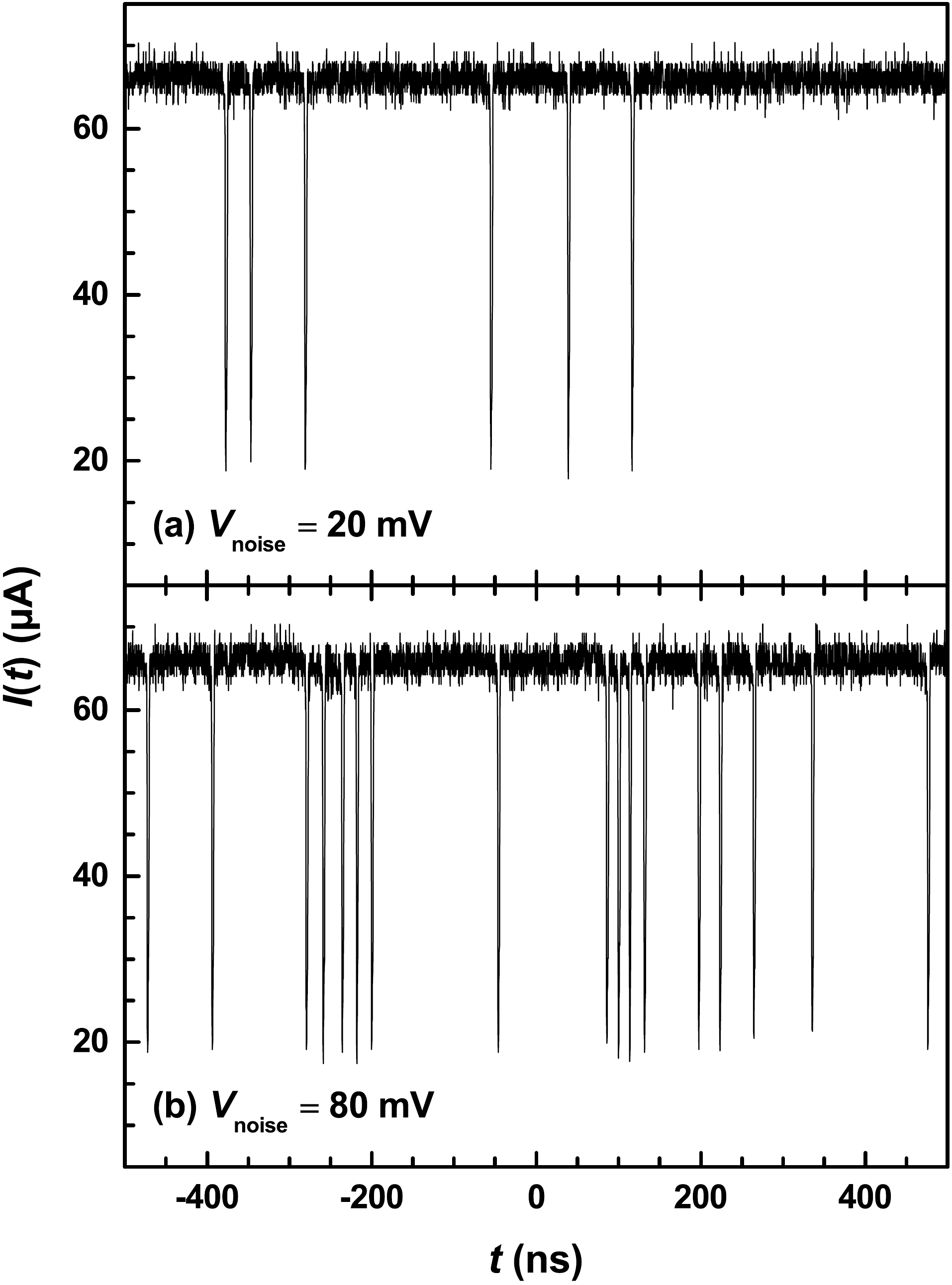}
\caption{Measured current oscillations $I(t)$ recorded for different noise amplitudes
(a) $V_\text{noise}=20$~mV and (b) $V_\text{noise}=80$~mV. $V_\text{DC}$ was fixed at $939$~mV.}
\label{Yin_Fig4}
\end{figure}
Current traces as a function of time have been recorded for two noise amplitudes at $V_\text{DC}=939$~mV
as shown in Fig.~\ref{Yin_Fig4}. Note that the current traces are recorded only
with AC coupling so that the information about the DC current level is lost.
For $V_\text{noise}=20$~mV, the current trace contains several irregularly
spaced spikes as shown in Fig.~\ref{Yin_Fig4}(a). When the noise amplitude is increased to $80$~mV,
the current trace exhibits more spikes with somewhat smaller amplitudes as shown in Fig.~\ref{Yin_Fig4}(b).
Similar current traces are observed for $V_\text{DC}=844$~mV (not shown). There are typically two oscillation
modes present in doped, weakly coupled SSLs, the dipole motion mode and the well-to-well hopping mode. When the
noise level is low, the dipoles are formed at the emitter moving toward the collector with a large interval
of noise spikes. As the noise amplitude is increased, a continuous motion of the well-to-well hopping of the
domain boundary occurs, introducing fast current oscillations in the SSL.

\begin{figure}[!b]
\includegraphics[width=0.45\textwidth]{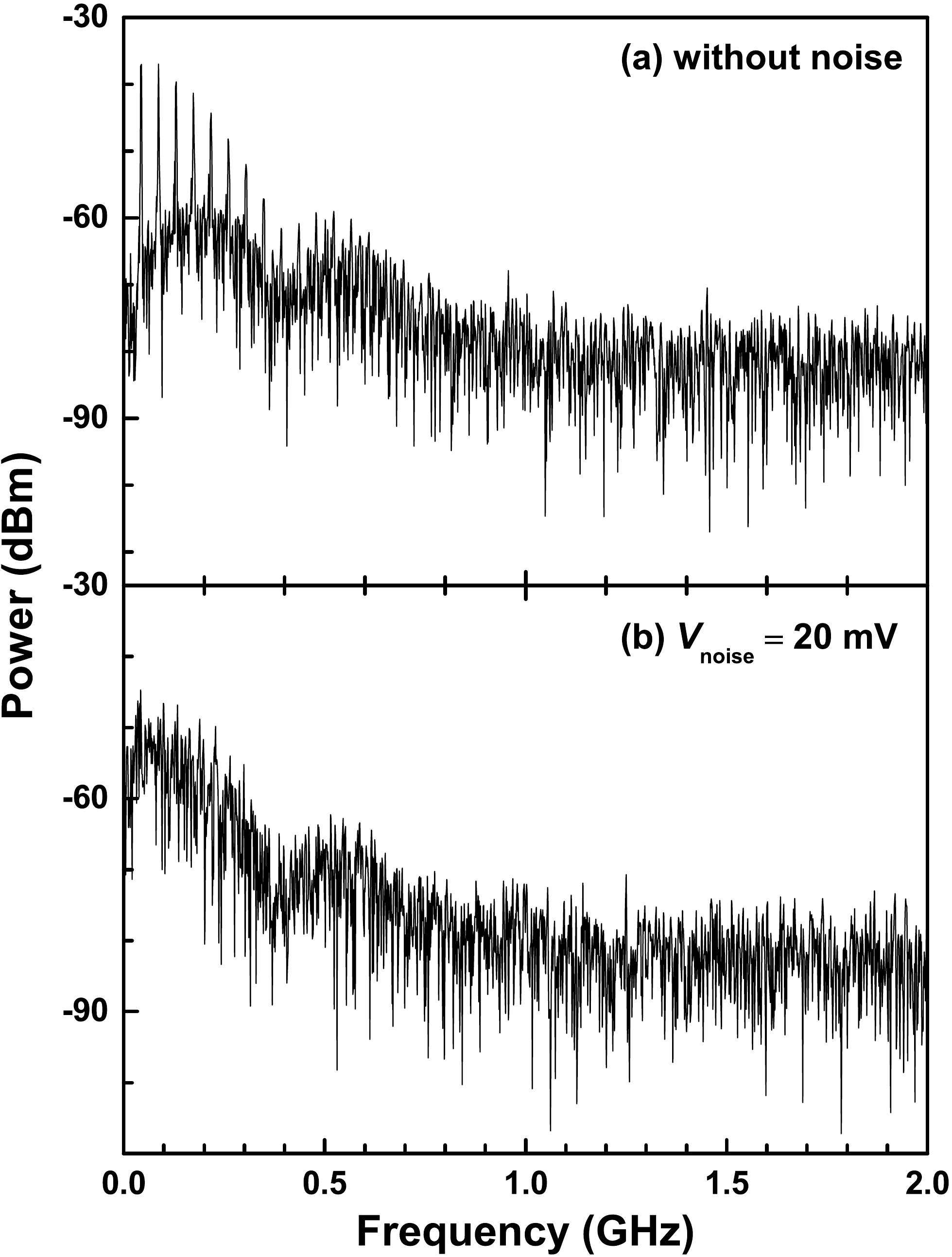}
\caption{Measured frequency spectra recorded (a) without any external noise and
(b) with $V_\text{noise}=20$~mV. $V_\text{DC}$ was fixed at $944$~mV.}
\label{Yin_Fig5}
\end{figure}
The power spectrum shown in Fig.~\ref{Yin_Fig5}(a) recorded at $V_\text{DC}=944$~mV
clearly demonstrates that, even without any external noise, the doped, weakly coupled SSL can already
exhibit a broad-band spectrum, indicating the possibility of the presence of chaotic
oscillations even without external noise. When noise is added as in Fig.~\ref{Yin_Fig5}(b),
the power spectrum becomes even broader. At the transition regions between periodic
oscillations and no oscillations, the system appears with increasing external noise amplitude
to exhibit a chaotic attractor, which leads to spontaneous chaos at room temperature.
As predicted by the results of the numerical simulations reported in Ref.~\onlinecite{alv14},
internal noise enhances spontaneous chaos. The same effect of chaos enhancement is observed
when we add external noise with increasing amplitude as will be further discussed below.
At the same time, the frequency spectra in the periodic region are dominated by an
oscillation with a single frequency, but with increasing bandwidth as the noise amplitude is increased.
Noise can significantly enhance the chaos generated in doped, weakly coupled SSLs.

\begin{figure}[!t]
\includegraphics[width=0.60\textwidth]{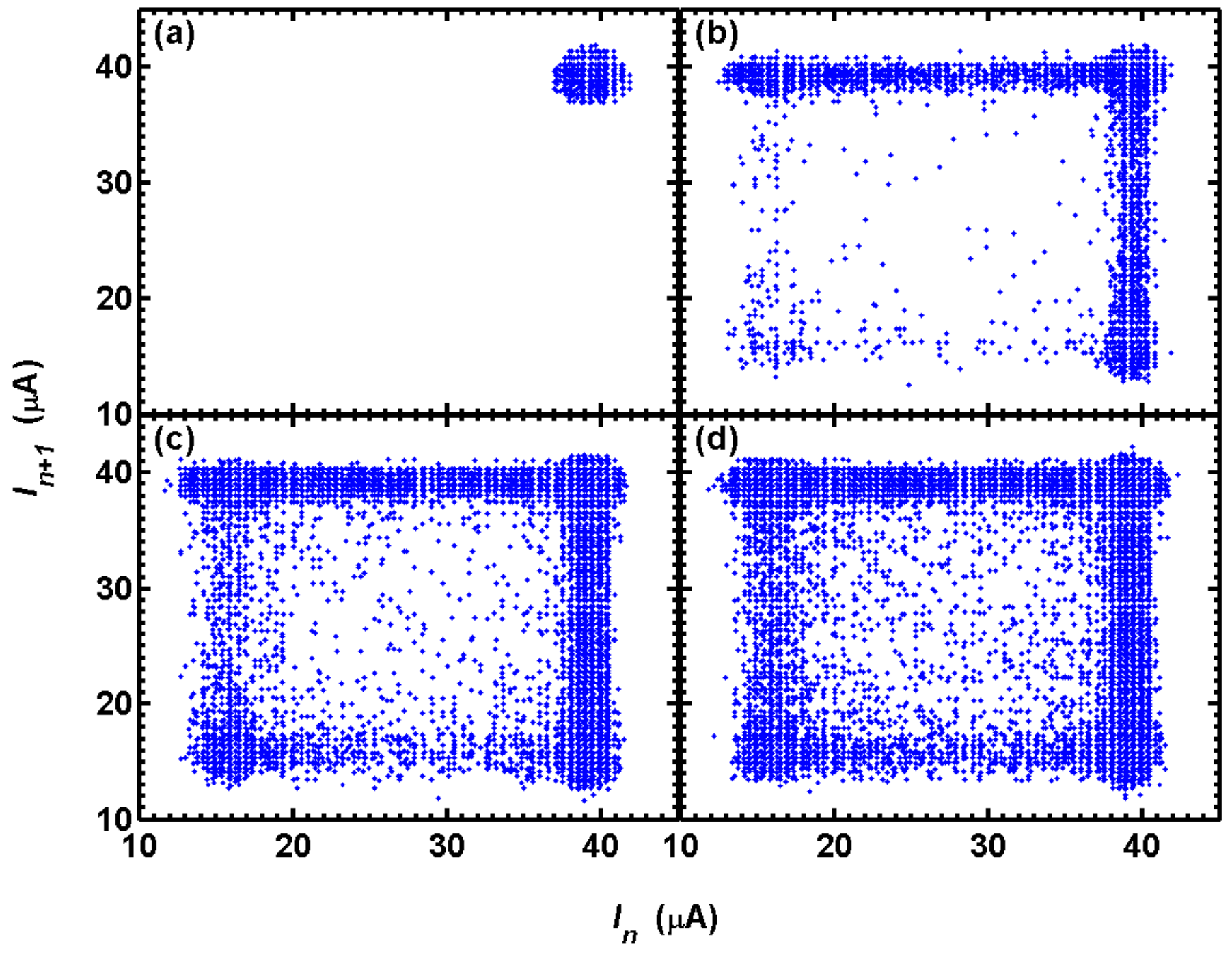}\\
\vspace*{1cm}
\includegraphics[width=0.60\textwidth]{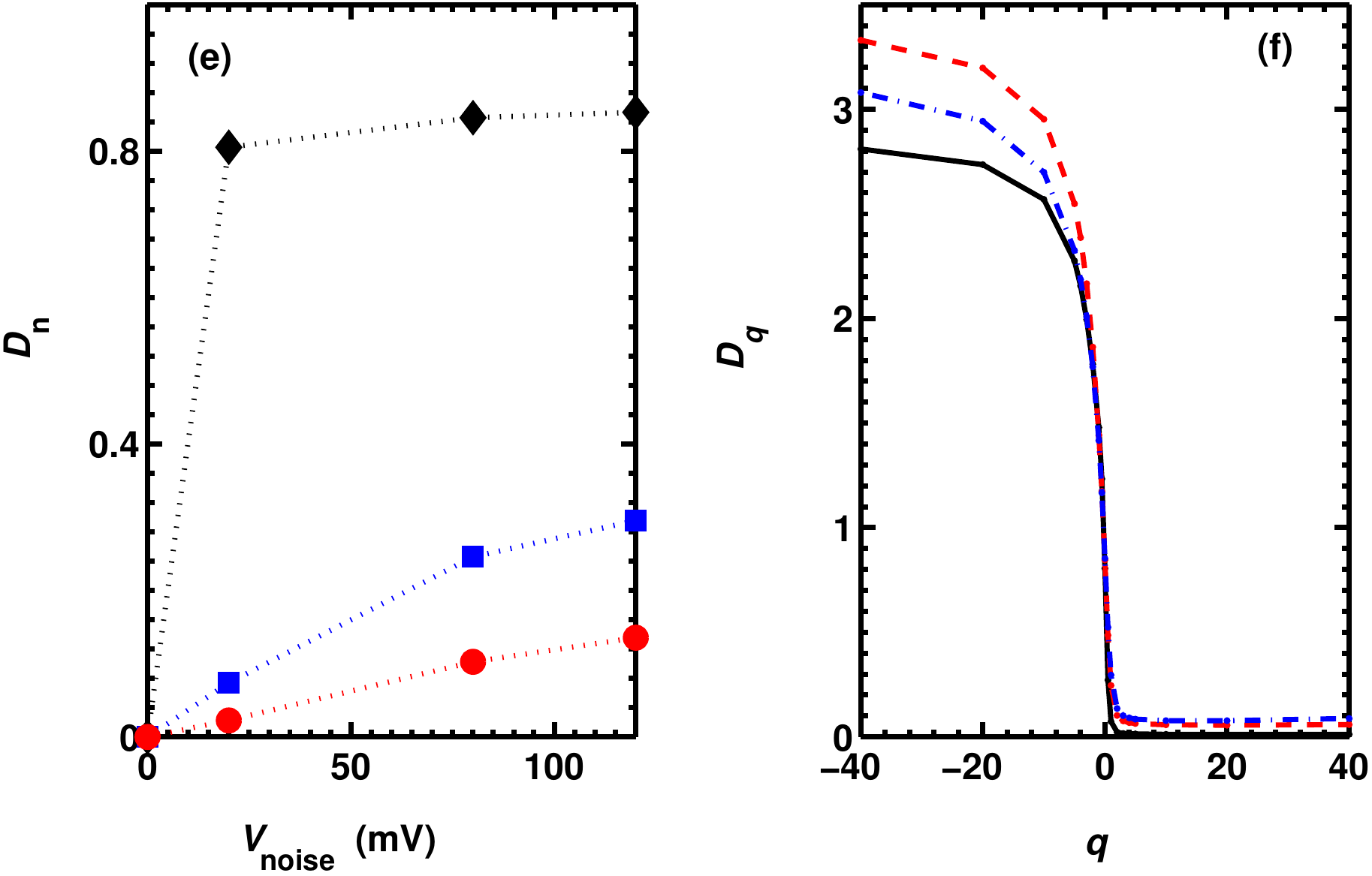}
\caption{Poincar\'e maps extracted from experimental current traces versus time
for (a) no external noise as well as with external noise of amplitude
(b) $V_\text{noise}=20$, (c) $80$, and (d) $120$~mV.
(e) Values of the capacity dimension $D_0$ (diamonds), the information dimension
$D_1$ (squares), and the correlation dimension $D_2$ (circles) as
a function of $V_\text{noise}$. (f) Multifractal spectra. $V_\text{noise} = 20$ (solid line),
$80$ (dashed line) and $120$ (dash-dotted line).  $V_\text{DC}$ was fixed at $845$~mV.}
\label{Yin_Fig6}
\end{figure}
\subsection{Attractors obtained from experimental results}
Figures~\ref{Yin_Fig6}(a), \ref{Yin_Fig6}(b), \ref{Yin_Fig6}(c), and \ref{Yin_Fig6}(d) show
reconstructed Poincar\'e maps from the experimental current traces versus time for no noise as
well as $V_\text{noise}=20$, $80$, and $120$~mV, respectively, at $V_\text{DC}=845$~mV.
In the absence of external noise, the deterministic attractor is stationary, and the
Poincar\'e map should consist of a single point. In Fig.~\ref{Yin_Fig6}(a), the unavoidable
internal noise produces many points concentrated in a small region. When external noise is
added, the Poincar\'e maps explore a region that becomes larger and fuller as the external
noise amplitude increases as shown in Figs.~\ref{Yin_Fig6}(b), \ref{Yin_Fig6}(c), and
\ref{Yin_Fig6}(d). Similar Poincar\'e maps are extracted from experimental current traces
versus time at $V_\text{DC}=940$~mV. Figure~\ref{Yin_Fig6}(e) shows the values of the
capacity dimension $D_0$, the information dimension $D_1$, and the correlation
dimension $D_2$ as a function of the external noise amplitude, which all increase with
increasing noise amplitude. Without external noise, the SSL is in a stationary state with
zero fractal dimension. When the external noise is turned on and increased, time-dependent
oscillations appear, and the fractal dimension $D_0$ increases abruptly demonstrating evidence
of chaotic oscillations induced by noise. Figure~\ref{Yin_Fig6}(f) shows the multifractal dimension
for the different noise amplitudes. For large positive values of $q$, the fractal dimension
$D_q$ corresponds to regions of the attractor that are more often visited by the system trajectory
and does not increase that much with external noise. For $q<0$ and $|q|$ large, $D_q$ is due
to the least sampled regions of the attractor, and it increases more strongly when the external
noise increases. We conclude that increasing the external noise expands the regions in
phase space that are visited by the system trajectory, an effect that is obvious from the
sequence shown in Figs.~\ref{Yin_Fig6}(b)--\ref{Yin_Fig6}(d).

\section{Model and simulated results \label{SecIII}}
\subsection{Discrete resonant tunneling model}
Numerical results are obtained using a discrete resonant tunneling model that captures
the main features of doped, weakly coupled SSL~\cite{bon02,bon05}. The model consists
of dynamical equations describing the evolution of the variables such as the electric
field $-F_i$ as well as the two-dimensional (2D) electron density $n_i$ at well $i$,
the tunneling current density $J_{i\rightarrow i+1}$ from well $i$ to $i+1$, and
the SL total current density $J(t)$:
\begin{equation}
\epsilon \frac{dF_i}{dt}+J_{i\rightarrow i+1}=J(t),
\end{equation}
\begin{equation}
J_{i\rightarrow i+1}=\frac{e n_i}{l} v^{(f)}(F_i)-J_{i\rightarrow i+1}^{-}(F_i,n_{i+1},T),
\end{equation}
\begin{equation}
J_{i\rightarrow i+1}^{-}(F_i,n_{i+1},T)=\frac{em^* k_B T}{\pi \hbar^2 l}v^{(f)}(F_i)
\ln \left[1+ e^{-\frac{eF_i l}{k_B T}}\left( e^{\frac{\pi \hbar^2 n_{i+1}}{m^* k_B T}} -1\right)  \right],
\end{equation}
\begin{equation}
n_i=N_D +\frac{\epsilon}{e}(F_i-F_{i-1}),
\end{equation}
\begin{equation}
\sum_{i=1}^{N}F_i=\frac{V_\text{DC}+\eta(t)}{l}, \quad \eta(t)=\eta_{th}(t)+\eta_{c}(t),
\end{equation}
\begin{equation}
J_{0 \rightarrow 1}=\sigma_0 F_0, \quad J_{N \rightarrow N+1} =\sigma_0 \frac{n_N}{N_D}F_N.
\end{equation}
Here $i=1,...,N$, where $N$ denotes the number of periods of the SSL. The forward electron velocity
$v^{(f)}(F_i)$ is a function with peaks corresponding to the discrete energy levels in every well
calculated using a Kronig-Penney model for the investigated SSL configuration as summarized
in Tab.~\ref{Yin_Tab1}. A more detailed description can be found in Refs.~\onlinecite{bon05}
and \onlinecite{alv14}. The overall voltage drop between the ends of the SSL $\sum_{i=1}^N F_i/l$
is equal to the DC voltage bias $V_\text{DC}$, while the noise voltage amplitude $\eta(t)$ corresponds
to $V_\text{noise}$, both provided in the experiment by the function generator in Fig.~\ref{Yin_Fig1}.
The numerical values of the parameters correspond to the
experimental configuration described in Sec.~\ref{SecII} with an equivalent 2D doping density due to the doping of the
central part of the quantum well of $N_D = 6$$\times$$10^{10}$~cm$^{-2}$. The effective mass of the
electrons in the GaAs/Al$_{x}$Ga$_{1-x}$As SL is $m^*= (0.063 +0.083x) m_e$, where
$m_e$ denotes the free-electron mass. The  well and barrier widths are $l_w =7$ and $l_b=4$ nm,
respectively, as for the experimentally investigated SSL so that the period of the SSL is
$l=l_b+l_w=11$~nm. The transversal area of the SSL is assumed to be $A=s^2$ with $s=30~\mu$m.
Finally, the relative permittivity of the SSL is $\epsilon= l/[l_w/\epsilon_w + l_b/\epsilon_b]$ with
$\epsilon_w=12.9\epsilon_0$ and $\epsilon_b=10.9\epsilon_0$ referring to the relative permittivity
of the well and barrier material, respectively. $\epsilon_0$, $-e<0$, $k_B$, $T$, and $\sigma_0$
denote the vacuum permittivity, the electron charge, Boltzmann's constant, the lattice temperature,
and the contact conductivity, respectively. The contact conductivity $\sigma_0$
is derived from a linear approximation of the emitter current density $J_{0\rightarrow 1}$,
which depends on the structure of the contact. We use $\sigma_0=0.763$~A/(V~m), which qualitatively
reproduces the experimental current-voltage characteristic shown in Fig.~\ref{Yin_Fig2}
for a barrier height $V_\text{barr}=388$~meV. The voltage noise $\eta(t)$ has two components:
(i) $\eta_{th}(t)$, which is related to the intrinsic noise of the source, and (ii) the
external noise $\eta_{c}(t)$. The thermal noise $\eta_{th}(t)$ is simulated by picking random
numbers every $5$$\times$$10^{-11}$~s from a zero mean distribution with a standard deviation of
$1.6$$\times$$10^{-3}$~V. The external noise $\eta_{c}(t)$ is
simulated by picking random numbers every $5$$\times$$10^{-9}$~s from a zero mean distribution
with a tunable standard deviation.
\begin{table}[!t]
\centering
\caption{Values of the potential barrier $V_\text{barr}$ and the first three energy levels $E_1$,
$E_2$, and $E_3$ for the doped, weakly coupled GaAs/Al$_{0.45}$Ga$_{0.55}$As SL.}
\label{Yin_Tab1}
\begin{tabular}{cccc}
\hline \hline
~~$V_\text{barr}$ (meV)~~& ~~$E_1$ (meV)~~& ~~$E_2$ (meV)~~& ~~$E_3$ (meV)~~\\ \hline
$388$ &	$45$& $173$ & $346$\\ \hline \hline
\end{tabular}
\end{table}

\subsection{Results of numerical simulations}
\begin{figure}[!t]
\includegraphics[width=0.45\textwidth]{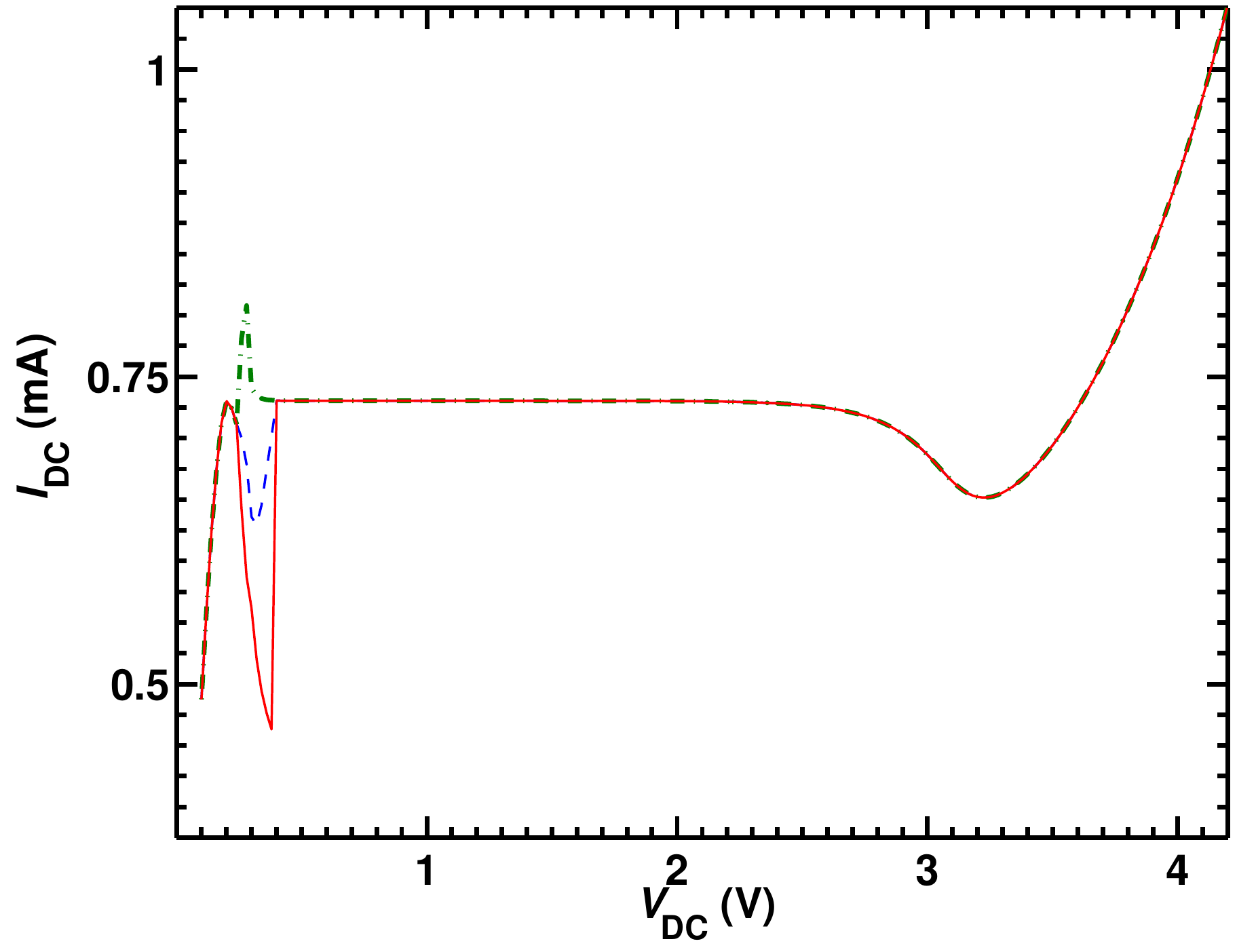}
\caption{Results of the numerical simulations of the current $I_\text{DC}$ versus voltage $V_\text{DC}$
for $\sigma=0.763$~A/(V~m). The dash-dotted, the dashed, and the solid line correspond
to the maximum, mean, and  minimum current, respectively.
In the voltage region between $0.26$ and $0.4$~V, current oscillations occur. The simulated
curves are shifted to lower voltages compared with the experimental curve due to built-in fields
and a voltage drop at an external resistance. Nevertheless, the oscillatory region is qualitatively
similar to the one found in the experiments as shown in Fig.~\ref{Yin_Fig2}.}
\label{Yin_Fig7}
\end{figure}

Figure~\ref{Yin_Fig7} depicts the simulated current-voltage characteristic for the deterministic
equations of our model [$\eta(t)=0$)]. The mean current (dashed line) should be compared with the
experimental one in Fig.~\ref{Yin_Fig2}. The average current for voltages at which there are
current oscillations in Fig.~\ref{Yin_Fig7} is qualitatively similar to the first part of
Fig.~\ref{Yin_Fig2}. There is a shift between the theoretical and experimental curves due to
the built-in voltage and a voltage drop at an external resistance that are absent in our model
equations. In the experiment, there is a voltage region of stationary current between two
regions of current oscillations. This window of stationary currents is absent in the
simulations of our model. However, the end of the first voltage region with current
oscillations in Fig.~\ref{Yin_Fig2} is similar to the end of the voltage region with
current oscillations in Fig.~\ref{Yin_Fig7}.

\begin{figure}
\includegraphics[width=0.60\textwidth]{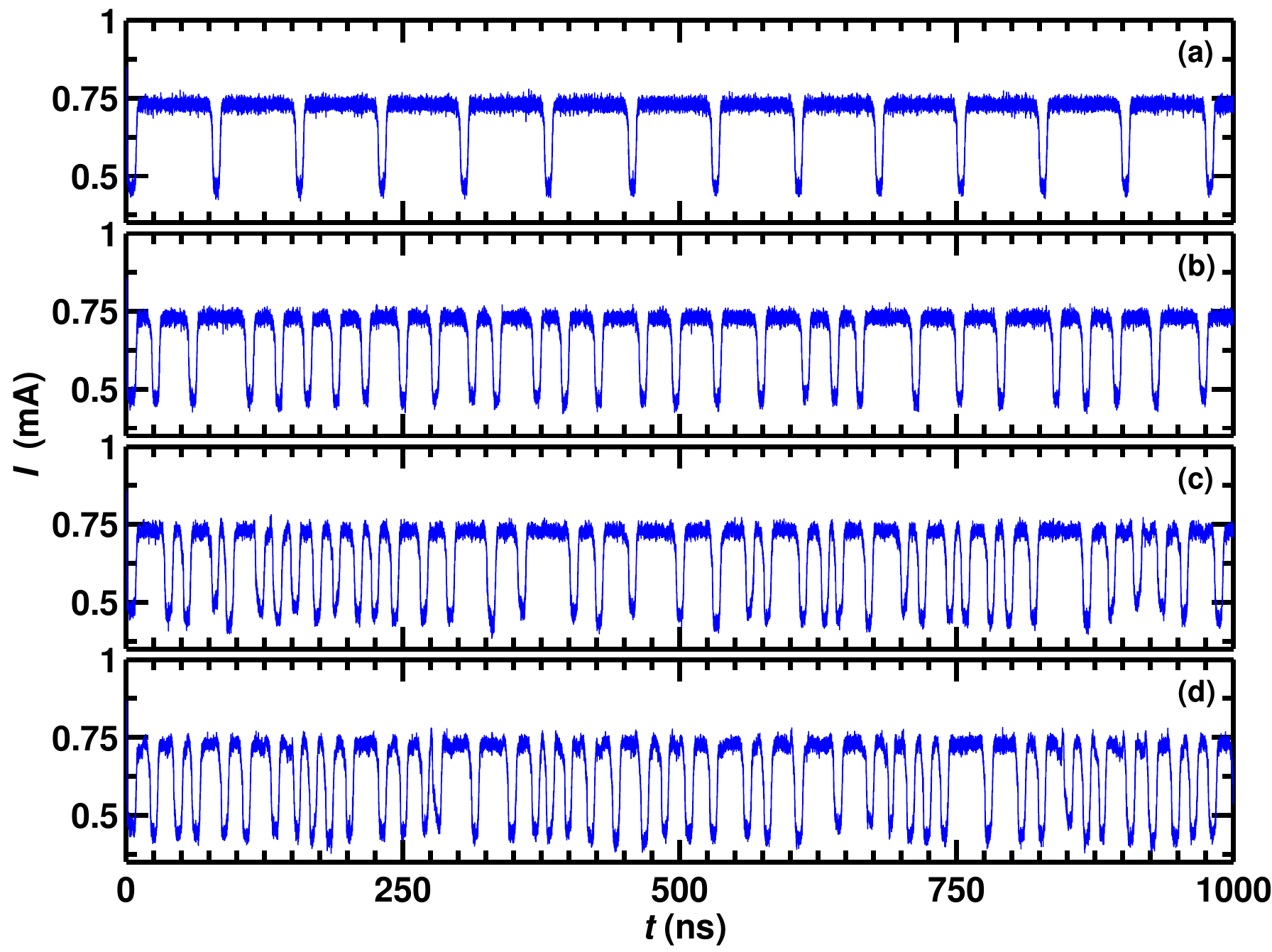}\\
\includegraphics[width=0.585\textwidth]{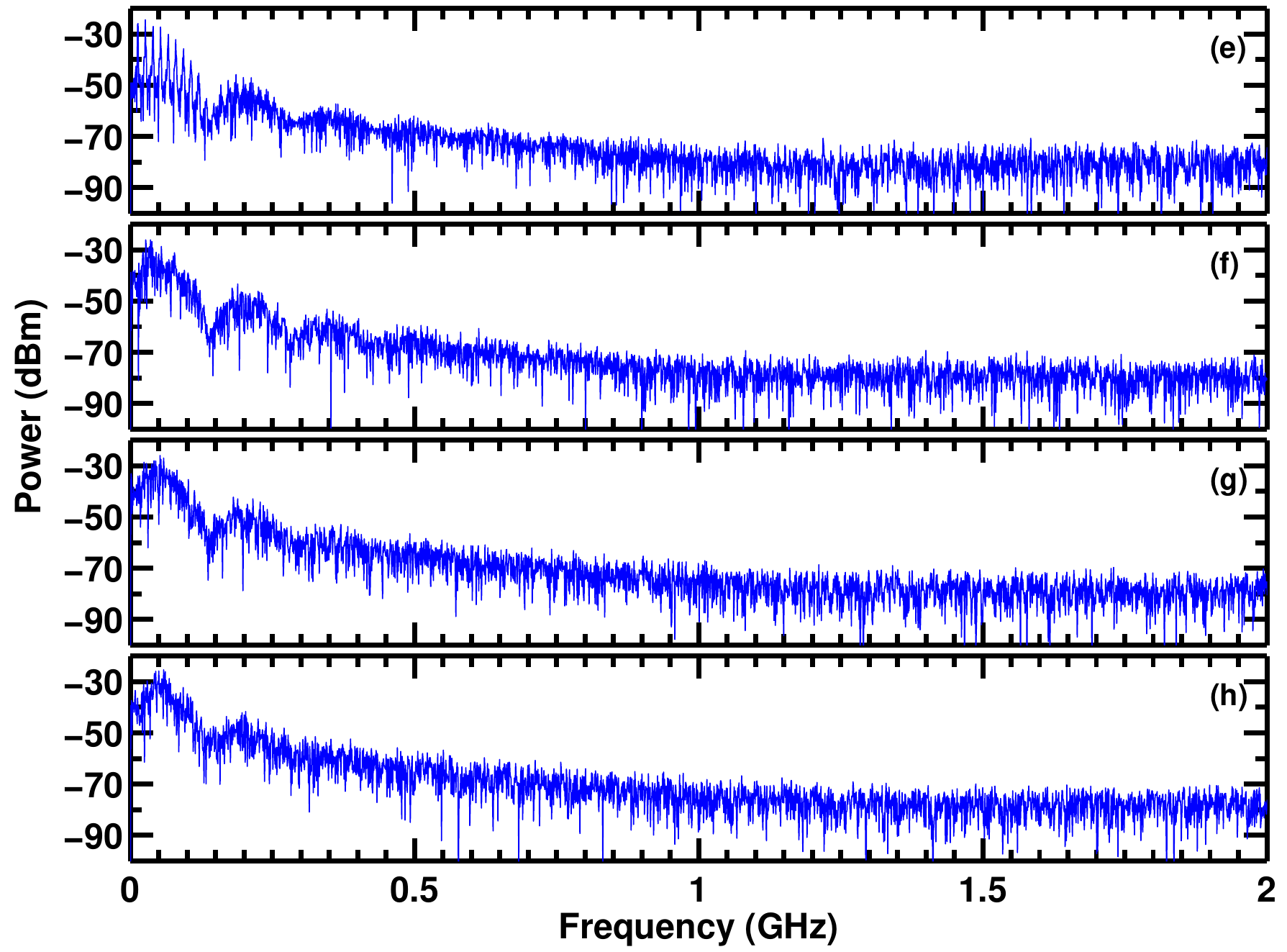}
\caption{Results of the numerical simulations for the current oscillations $I(t)$ for different
noise amplitude (a) $V_\text{noise}=1$, (b) $8$, (c) $33$, and (d) $46$~mV at
$V_\text{DC}=380$~mV in the oscillatory region of the current-voltage
characteristics presented in
Fig.~\ref{Yin_Fig7}. The corresponding frequency spectra as determined by a
numerical Fourier transform are shown in (e), (f), (g), and (h), respectively.}
\label{Yin_Fig8}
\end{figure}

Figures~\ref{Yin_Fig8}(a), \ref{Yin_Fig8}(b), \ref{Yin_Fig8}(c), and \ref{Yin_Fig8}(d)
show the results of the numerical simulations for the current oscillations for different
noise amplitudes (a) $V_\text{noise}=1$ (only thermal noise is present), (b) $8$, (c) $33$, and (d) $46$~mV, respectively,
at $V_\text{DC}=380$~mV, which lies in the oscillatory region of the current-voltage curve presented in
Fig.~\ref{Yin_Fig7}. The internal and external noise are numerically generated as explained above.
With increasing $V_\text{noise}$, the oscillations become less periodic,
contain more spikes in the same time interval, and the amplitude becomes more and more
irregular. The corresponding frequency spectra as determined by a
numerical Fourier transform are displayed in Fig.~\ref{Yin_Fig8}(e), \ref{Yin_Fig8}(f),
\ref{Yin_Fig8}(g), and \ref{Yin_Fig8}(h), respectively.
The number of peaks in the frequency spectra
decreases with increasing noise amplitude, while the width of the frequency spectra increases.
We have repeated the simulations for a voltage of $460$~mV, for which the deterministic system
is in a stationary state. The behavior with increasing noise amplitude is very similar to the
one shown in Figs.~\ref{Yin_Fig8}(b)--\ref{Yin_Fig8}(d) with some differences in the details of
the current oscillations and frequency spectra. In comparison to Fig.~\ref{Yin_Fig8}(a), it looks
of course different, because there are no current oscillations in this case. Overall, the experimental
trend with increasing noise amplitude is reproduced, i.~e., the larger the controlled noise amplitude,
the more random the current oscillations become. Note that simulations of the field
distribution (not shown) indicate that the current spikes in the simulated current traces in
Figs.~\ref{Yin_Fig8}(a)--\ref{Yin_Fig8}(d) are caused by the generation of a small charge dipole,
which is generated at the emitter contact and moves toward the collector contact. When the front part
of the dipole reaches the collector contact, it disappears, while the back part of the dipole rebuilds
the domain wall, which existed before the current spike.

\begin{figure}[!b]
\includegraphics[width=0.70\textwidth]{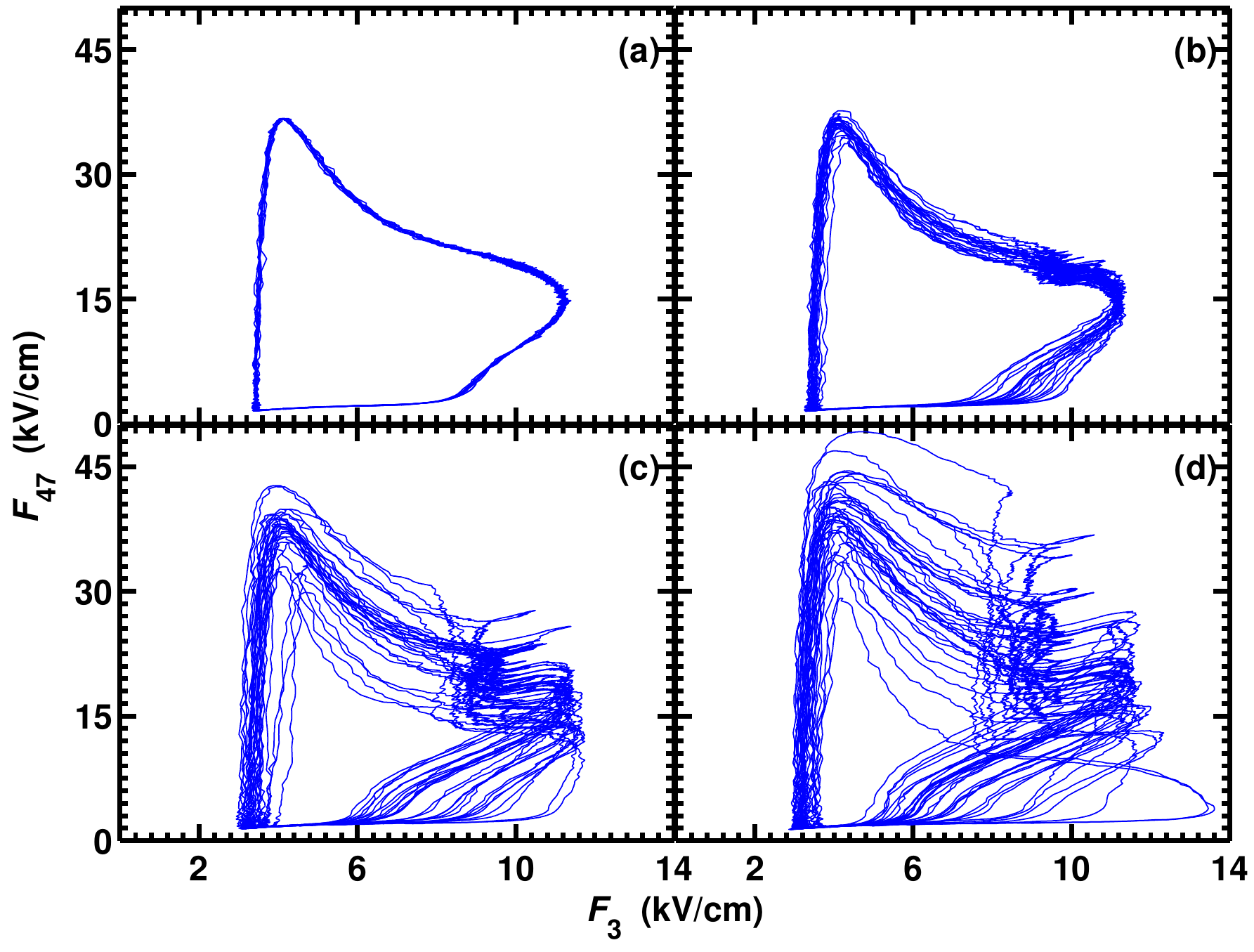}
\caption{Results of the numerical simulations for the phase portraits, where the electric field
in well number $47$ is presented versus the electric field in well number $3$.
The noise amplitude increases from (a) $V_\text{noise}=1$, (b) $8$, (c) $33$, to
(d) $46$~mV at $V_\text{DC}=380$~mV in the oscillatory region of the current-voltage
characteristics presented in Fig.~\ref{Yin_Fig7}. Loops correspond to
oscillations depicted in Fig.~\ref{Yin_Fig8}.}
\label{Yin_Fig9}
\end{figure}
We have also constructed phase portraits from the calculated electric field in well
number $47$ versus well number $3$. We can build alternative Poincar\'e maps based on the
intersection of the trajectories in these phase portraits with an appropriate segment of a
straight line. Period doubling scenarios or quasi-periodic attractors could be visualized
using such Poincar\'e maps. Thus, these phase portraits give information on the chaotic state of the system.
Figures~\ref{Yin_Fig9}(a), \ref{Yin_Fig9}(b), \ref{Yin_Fig9}(c), and \ref{Yin_Fig9}(d) show the
phase portraits for different noise amplitude (a) $V_\text{noise}=1$, (b) $8$, (c) $33$, and (d) $46$~mV,
respectively, at $V_\text{DC}=380$~mV, which lies in the oscillatory region of the current-voltage
curve presented in Fig.~\ref{Yin_Fig7} and corresponds to the simulated data in Fig.~\ref{Yin_Fig8}.
When only \emph{thermal} noise is present [cf. Fig.~\ref{Yin_Fig9}(a)], the oscillations are periodic,
even though they are somewhat distorted by noise. However, when controlled noise is
added as shown in Figs.~\ref{Yin_Fig9}(b)--\ref{Yin_Fig9}(d), the oscillations loose their
periodicity and become more and more distorted with increasing noise amplitude, i.~e.
the complexity of the phase portraits strongly increases, indicating increasing
chaotic behavior. The phase portraits for a voltage of $460$~mV (not shown), for which the deterministic
system is in a stationary state, look qualitatively very similar to the ones shown in
Fig.~\ref{Yin_Fig9}, except for the first one with only \emph{thermal} noise, because there are no
current oscillations in this case. Note that the \emph{thermal} noise does not distort
the periodicity, since it only introduces some small fluctuations. However, by
introducing a longer time-correlated (\emph{controlled}) noise, clear chaotic
behavior is obtained.

\section{Summary and conclusions \label{SecIV}}
We have observed that the external noise can induce spontaneous chaotic oscillation
in a narrow voltage interval in a doped, weakly coupled GaAs/(Al,Ga)As SL at room temperature.
Results of numerical simulations of nonlinear transport based on a discrete tunneling model
qualitatively confirm the experimentally observed features. While in a noise-free SSL static
domain boundaries are formed, the domain boundary moves into the collector, new domain
walls are formed near the emitter, and chaotic current
oscillations are induced, when noise is added. Therefore, with increasing noise amplitude,
chaotic current oscillations can be enhanced in semiconductor superlattices. This approach
is consequently a very robust method of producing chaotic current oscillations in
doped, weakly coupled semiconductor superlattices.

\section*{Acknowledgements}
The authors would like to thank the Strategic Leading Science and Technology Special
of the Chinese Academy of Sciences (grant XDA06010705), the National Natural Science
Foundation of China (grant 61204093) and the Ministerio de Econom\'\i a y Competitividad
of Spain (grant MTM2014-56948-C2-2-P) for financial support.

\end{document}